\documentclass[11pt,letterpaper]{article}

\usepackage[utf8]{inputenc}

\usepackage{latexsym}
\usepackage{verbatim}
\usepackage[a4paper, total={6.5in, 9in}]{geometry}
\usepackage{silence}
\usepackage{amsthm,amsmath,amsfonts,amssymb,verbatim, color}
\usepackage{graphicx}
\usepackage[percent]{overpic}
\usepackage{bm}
\usepackage{braket}
\usepackage{float}
\usepackage{mathtools}
\usepackage{appendix}
\usepackage[colorlinks=true,citecolor=blue,linkcolor=black,urlcolor=blue]{hyperref}
\usepackage{graphics} % Testing...
\graphicspath{ {figures/} }
\usepackage[caption=false]{subfig}

\usepackage[mathscr]{euscript}
\usepackage[export]{adjustbox}
\usepackage[bottom]{footmisc}
\usepackage{textcomp}
\usepackage{braket}
\begin{document}

\title{A Genus-two Surface Code}
\author{$^{1,2}$Komal Kumari, $^{1}$Garima Rajpoot, and $^{1,2}$Sudhir Ranjan Jain \\ $^{1}${\small{\it Theoretical Nuclear Physics and Quantum Computing Section}}\\ {\small{\it Nuclear Physics Division, Bhabha Atomic Research Centre, Mumbai 400085, India}}\\ $^{2}${\small{\it Homi Bhabha National Institute, Training School Complex, Anushakti Nagar, Mumbai 400094, India}}
}
\maketitle
\begin{abstract}
We construct a genus-two surface code by exploiting the planar tessellation using a rhombus-shaped tile. With \textit{n} data qubits, we are able to encode at least {n/3} logical qubits or quantum memories. By a suitable arrangement of the tiles, the code achieves larger distances, leading to significant error-correcting capability. We demonstrate the robustness of the logical qubits thus obtained in the presence of external noise. We believe that the optimality of the code presented here will pave the way for design of efficient scalable architectures. 
\end{abstract}

\section{Introduction}
Computations and errors go hand in hand. It is well-known that quantum error correction  has paved the way for fault-tolerant computation \cite{Kitaev,preskill1998fault,aharonov2008fault}. The challenges posed by no-cloning theorem, existence of continuous errors, and projective measurements (collapse of state) have been overcome by the usage of entanglement \cite{krjj}, Hadamard basis, and syndrome decoding based on parity check matrices (PCM). This was realized by introducing the quantum stabilizer codes (QSC) \cite{gottesman1997stabilizer}, a profound generalization of classical coding \cite{hamming1986coding,huffman2010fundamentals}. Essentially, $[[n, k]]$ QSC maps the state carrying information (logical qubit) in $2^k$-dimensional Hilbert space onto a codeword in $2^n$-dimensional Hilbert space. The redundancy in superpositions of the entangled states and measurement of the ancillae allows one to extract information from the system without knowing the state of individual qubits. Perhaps the most useful approach to achieve fault-tolerance is via developing a surface code \cite{bravyi1998quantum,fowler2012surface}. It is a topological code  where the principle of building the code is to ``patch" together repeated elements. It consists of a lattice of squares with alternating physical qubits acting as data qubits and ancillary qubits, the latter measuring $X$ and $Z$.
\begin{figure*}[ht!]    
    \centering
    \subfloat[]{\includegraphics[width=0.37\textwidth]{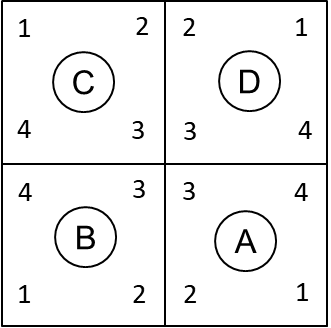}}\hspace{15mm}
    \subfloat[]{\includegraphics[width=0.4\textwidth]{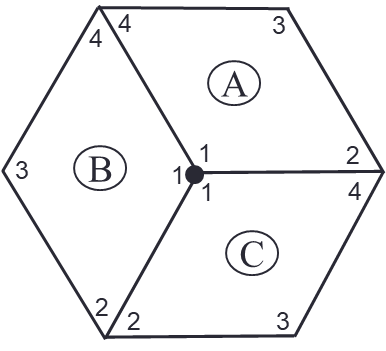}}  
    \caption{(a) Four subsequent reflections of a square about its edges reveal that  orientation is preserved as the (numbered) vertices return to the original square. The two vertical edges (1441) are identified to form a cylinder. Further identification of the edges (1221) gives the shape of a torus. On the other hand, (b) in the $(\pi/3, 2\pi/3)$-rhombus shape, upon three successive reflections about its sides returns to a flipped configuration. It takes three more to restore the configuration, albeit at the expense of losing single-valuedness. Thus, the vertex marked with a black dot is like a branch point. Surface corresponding to this shape is shown in Fig. \ref{fig:gistwo}. Tessellation with this shape leads to visiting certain straight segments twice - these form branch cuts (see Fig. \ref{fig:G}).}
\label{fig:reflection}
\end{figure*}

Planarity helps in realizing an efficient engineering design of a chip accommodating qubits and ancillae. This modular approach leads to scalability while ensuring stabilizer commutativity. Surface code requires only  nearest-neighbour interaction which facilitates  implementation on hardware. While this code (a) is provably fault-tolerant under general noise models \cite{chubb2021statistical}, (b) supports efficient decoding algorithms \cite{hutter2014efficient} and a fault-tolerant implementation of gates  \cite{bombin2007topological, horsman2012surface}, it requires a very large overhead in the number of physical qubits \cite{fowler2012surface}. It was shown experimentally \cite{tomita2014low} that in surface code of $9$ data qubits, the decoherence time of encoded logical qubits is much higher than that of physical qubits. For lesser overhead, using Hadamard rotation on alternate data qubits and syndrome measurement given by the product $XZZX$ on each qubit, an efficient fault-tolerant variant of the surface code \cite{bonilla2021xzzx} was presented. Further improvement in the overhead was achieved by the topological stabilizer code, $XYZ^2$ on a honeycomb structure \cite{srivastava2022xyz}. Both of these codes have better performance for biased noise. 
\begin{figure}[htbp!]
    \begin{center}
    \includegraphics[width=0.95\textwidth]{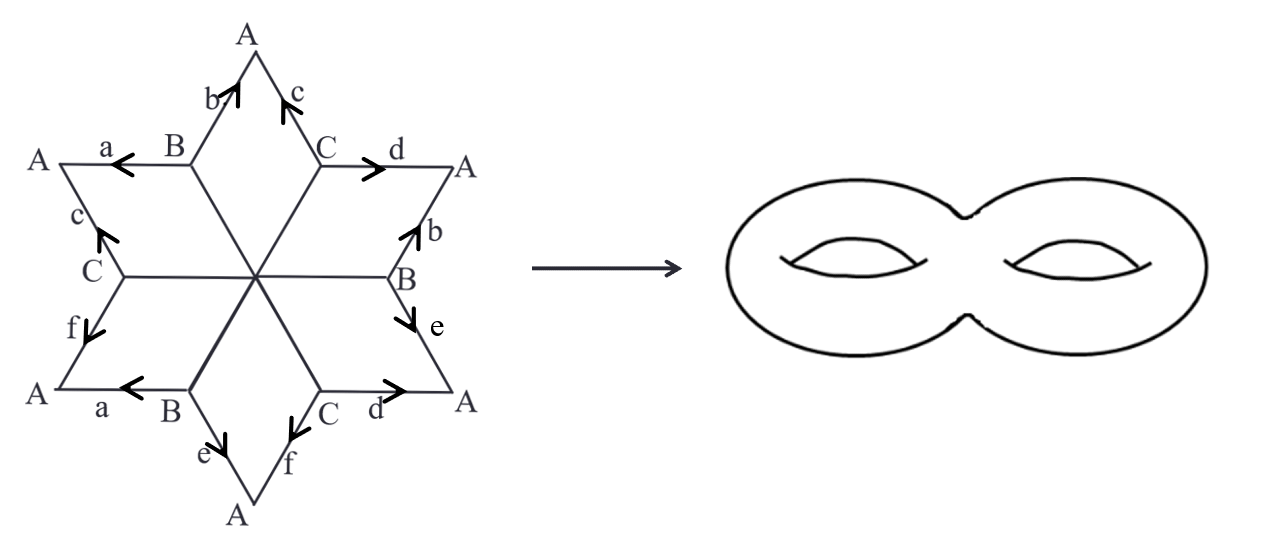}
    \caption{\small{Six copies of the rhombus are required to construct an invariant surface. By identifying the edges labelled by the same arrows, the shape shown folds into a surface topologically equivalent to a sphere with two handles (genus two).}}
    \label{fig:gistwo}
    \end{center}
\end{figure}

The most striking consequence of this design is to look for ``unit shapes" that can tessellate the plane upon successive reflections. Based on seminal work by Weyl \cite{weyl1926nachtrag} and Cartan \cite{cartan1927geometrie}, Coxeter \cite{coxeter1973regular} pointed out a one-to-one correspondence between Lie groups and reflection groups whose fundamental regions are simplexes in Euclidean space. These fundamental regions generate tori for ``unit shapes" like square, equilateral triangle, right isosceles triangle, or a hemi-equilateral triangle \cite{jain2017nodal}. To illustrate, if we consider a square tile, upon successive reflections about its sides, it can be easily seen that four copies, making a larger square with an edge-length twice the length of the original square, forms a unit of tessellation - the fundamental domain, identifying the pairs of  parallel edges gives a torus, which is characterized by a topological invariant, the genus being equal to one.  Beautiful multiply-connected shapes form by identifying edges of rational triangles and rhombi in a fundamental domain in a certain way   \cite{eckhardt1984analytically,jain1992periodic}. An illustration of the association of $(\pi/3, 2\pi/3)$-rhombus and a double torus (genus two) is shown in Fig. \ref{fig:gistwo}.  

\begin{figure}[htbp!]
    \begin{center}
    \includegraphics[width=0.7\textwidth]{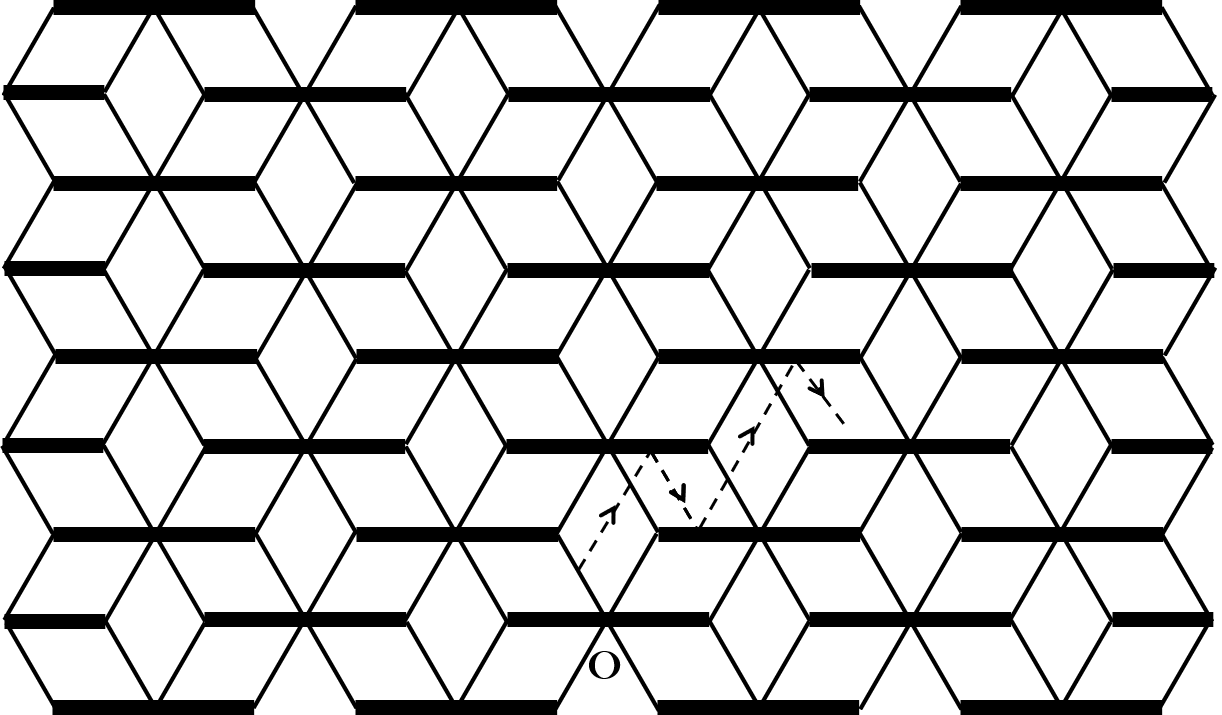}
    \caption{\small{A plane can be filled by successively reflecting a $\pi/3$-rhombus about its sides. The arrangement of the rhombi is shown here with bold segments representing the regions visited twice. Upon identification of corresponding sides, one obtains the surface of a sphere with two handles (see Fig. \ref{fig:gistwo}). To compare, with a square as a basic unit, one obtains the surface of a torus. See further details in Ref. \cite{jain2017nodal}. Also see Fig. \ref{fig:reflection}}}
    \label{fig:G}
    \end{center}
\end{figure}

Here we introduce a surface code corresponding to a surface with genus two - topologically equivalent to a sphere with two handles - obeying stabilizer algebra. The basic tile shape is a $(\pi/3, 2\pi/3)$-rhombus, the intricacy of this simple shape is illustrated in Fig. \ref{fig:reflection}. The fundamental domain is constructed by stitching six copies which, via an appropriate identification of edges (Fig. \ref{fig:gistwo}), creates a ``genus-two surface" (double torus) \cite{jain2017nodal,eckhardt1984analytically,richens1981pseudointegrable}, hence the name of the code. Filling of the plane is shown in Fig. \ref{fig:G} where the bold segments correspond to regions visited twice upon successive reflections. However, as shown below, a stabilizer code can be built nevertheless.

\begin{figure}[htbp!]
    \centering
    \includegraphics[width=0.25\textwidth]{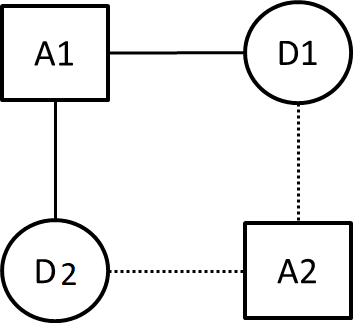}
    \caption{Figure shows the basic surface code-4 cycle which consists of an X-ancilla ($A1$), Z-ancilla ($A2$) and two data qubits ($Di$). Bold lines show the Control-$X$ and dotted lines show the Control-$Z$ operations on the data qubits.}
    \label{sc-4-cycle}
\end{figure}

\section{Four-cycle, [[5,1,2]] surface code: re-interpretation}

To begin with, we go back to the very beginning - the four-cycle (Fig. \ref{sc-4-cycle}) where we have two data qubits, each addressed by an $X$ and $Z$ ancilla \cite{fowler2012surface}. The bold and dashed lines represent Control-$X$ and Control-$Z$ operations to respectively detect $Z$ (dephasing) and $X$ (bit flip) errors. As this has two code qubits and two stabilizers ($X_1X_2$, $Z_1Z_2$), there cannot be any  logical qubit. Even if this is not useful, by tiling it, useful codes have been realized \cite{fowler2012surface}. 

\begin{figure*}[ht!]    
    \centering
    \subfloat[]{\includegraphics[width=0.4\textwidth]{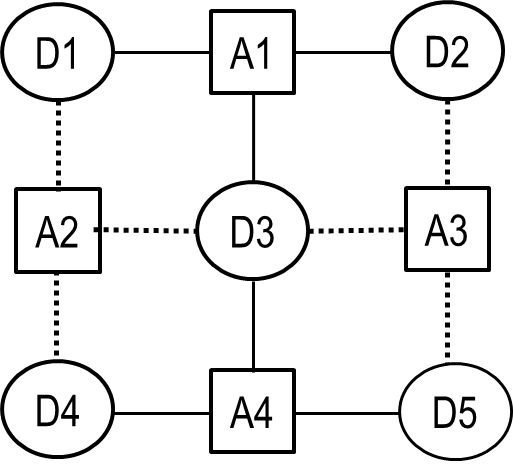}}\hspace{15mm}
    %\subfloat[]{\includegraphics[width=0.35\textwidth]{square_reflections.png}}  
    %\subfloat[]{\includegraphics[width=0.35\textwidth]{ent_al_1.7_x1-x7_0.4.png}}
    %\subfigure[]{\includegraphics[width=0.24\textwidth]{monalisa.jpg}}
    \caption{In [[5,1,2]] surface code shown here, $Ai$'s represent the ancillae and $Di$'s represent  the data qubits. Bold (dotted) lines depict Control-$X$ ($Z$) operations. The analogy between the two is compelling and suggestive. In both cases, upon identification of pair of sides, one obtains a torus.}
\label{512code}
\end{figure*}

The simplest construction, minimal in the sense of tiling (referring to Fig. \ref{fig:reflection}), gives the  $[[5,1,2]]$ surface code (Fig. \ref{512code}). We see that the stabilizers $P$ contain four commuting elements $\{X_1X_2X_3, Z_1Z_3Z_4, Z_2Z_3Z_5, X_3X_4X_5\}$. With five code qubits, there is one logical qubit. The pair of logical operators which obeys the stabilizer algebra is $\langle\bar{X}_1=X_1X_4$, $\bar{Z}_1=Z_1Z_2\rangle$. With the stabilizers and the logical operators, the two logical states of this code are \cite{gottesman1997stabilizer}

\begin{alignat}{1}
   \ket{0}_L&=\frac{1}{\mathcal{N}}\prod_{P_i\in\langle P\rangle}(I^{\otimes n}+P_i)\ket{0^{\otimes n}}\nonumber\\
   &= \frac{1}{{\mathcal N}}(\ket{00000}+\ket{11100}+\ket{00111}+\ket{11011})\nonumber\\
   \ket{1}_L&=X_1X_4\ket{0}_L=\frac{1}{{\mathcal N}} (\ket{10010}+\ket{01110}+\ket{10101}+\ket{01001}).
\end{alignat}
The tiling procedure can be continued, leading to a tessellation of the plane. 

We would like to observe that in Fig. \ref{fig:reflection}, if we start from square $A$ with vertices marked as shown, then upon four such reflections, we obtain the square $A$ back. Arranging data and ancilla qubits on vertices of this square as in Fig. \ref{512code}, the edges $D_1A_2D_4$ and $D_2A_3D_5$ may be identified to give a cylinder; subsequent identification of $D_1A_1D_2$ and $D_4A_4D_5$ will yield the shape of a torus \cite{footnote}.

In this paper, we consider this interpretation of tiling rather than its association with the Ising spin model. We consider different tiling shapes, inspired by non-integrable billiards, and show that a stabilizer code can indeed be built with much more efficient encryption. This is the central thesis of our work.

\section{Unit of the genus-two code}
\begin{figure*}[ht!]    
    \centering
    \subfloat[]{\includegraphics[width=0.5\textwidth]{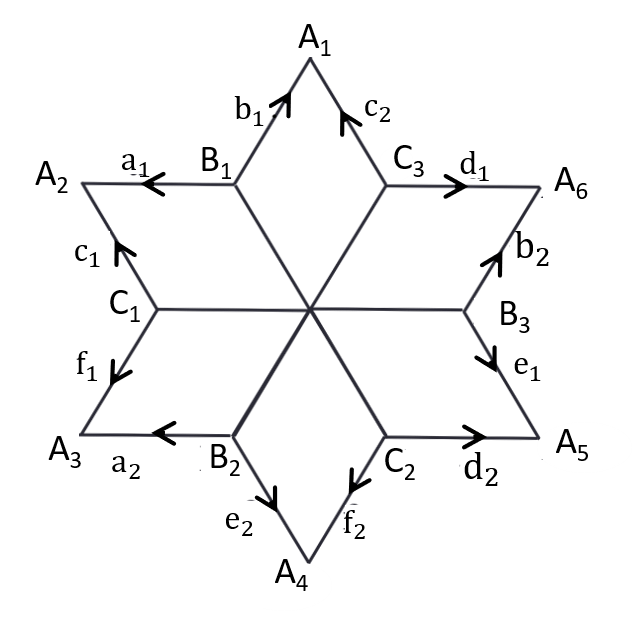}} 
    \subfloat[]{\includegraphics[width=0.45\textwidth]{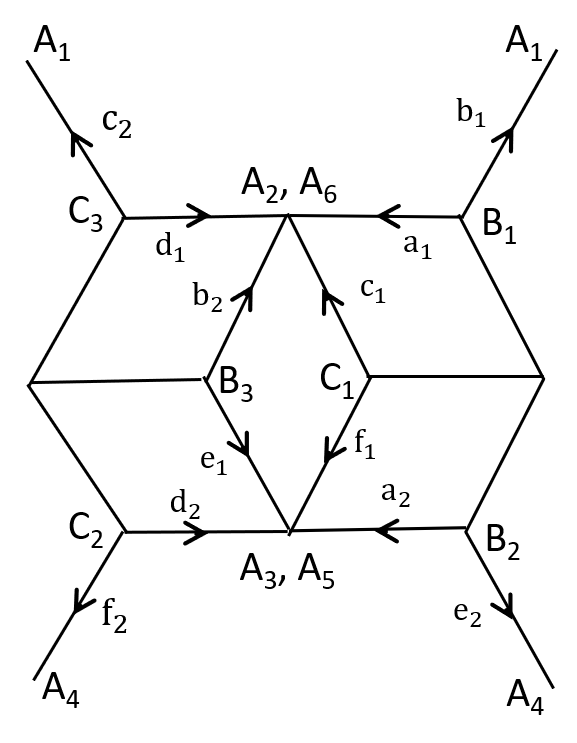}}\\
    \subfloat[]{\includegraphics[width=0.6\textwidth]{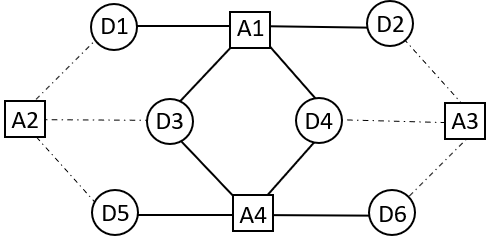}}
    \caption{The sides are marked as $a_1, a_2,\dots, f_1, f_2$ and the vertices are marked as $A_1,\dots, A_6$, $B_1, B_2, B_3$, $C_1,C_2$ and $C_3$. These arrows are used for identifying the edges in (a). Based upon these identifications, the equivalent structure is shown in (b). This structure will be used as the basic unit of the code. (c) The data qubits are represented by $Di$ and ancillary qubits by $Ai$. The solid (dashed) lines represent Control-$X$ (Control-$Z$)-operations on data qubits. The data and ancillary qubits are placed alternately in the spirit of \cite{fowler2012surface}.}
\label{fig:G2}
\end{figure*}

Let us begin with a simple illustration of the structure of the code where a unit consists of $10$ physical qubits, comprising of $n=6$ data ($D$) qubits ($m=4$ ancilla, ($A$)), represented by circles (squares),  see Fig. \ref{fig:G2}. The bold (dotted) lines represent $X$ ($Z$) - measurement. For instance, ancilla qubits $A1$ and $A4$ ($A2$ and $A3$) perform $X$ ($Z$) measurement on data qubits. 
States of data qubits are acted upon by the operators belonging to the Pauli group $\mathcal{G}_6$ \cite{footnote1}. We look for operators which would preserve logical qubit, $|\bar{\psi}\rangle$ i.e., satisfy $P_i|\bar{\psi}\rangle=(+1)\ket{\bar{\psi}}$. For this, a set of stabilizers is $P=$\{$X_1X_2X_3X_4$, $X_3X_4X_5X_6$, $Z_1Z_3Z_5$, $Z_2Z_4Z_6$\}. A logical $\ket{0}_L$ qubit of this code  is \cite{gottesman1997stabilizer}:
\begin{alignat}{1}\label{eq:lzero}
    \ket{0}_L&=\frac{1}{\mathcal{N}}\prod_{P_i\in\langle P\rangle}(I^{\otimes n}+P_i)\ket{0^{\otimes n}}\nonumber\\
    &=\frac{1}{\mathcal{N}}(I^{\otimes 6}+X_1X_2X_3X_4)(I^{\otimes 6}+X_3X_4X_5X_6)\nonumber\\
    &\hspace{9mm}(I^{\otimes 6}+Z_1Z_3Z_5)(I^{\otimes 6}+Z_2Z_4Z_6)|0^{\otimes 6}\rangle\nonumber\\
    &=\frac{1}{\mathcal{N}}(\ket{000000}+\ket{001111}+\ket{111100}+\ket{110011}),
\end{alignat}
where $\mathcal{N}$ is an overall normalization constant. As can be easily verified, all stabilizers commute with each other, i.e., $[P_i,P_j]=0, \forall$ $i,j$.  To transform $|0\rangle_L$ to $|1\rangle_L$, we would have to construct a logical-$X$ operator, and similarly, a logical-$Z$ operator. Such pairs of logical operators $\bar{X}$'s and $\bar{Z}$'s must 
(i) commute with all the stabilizers in $P$ 
(ii) anti-commute pairwise, i.e. $\{\bar{X}_{i}, \bar{Z}_{i}\} = 0\, \forall \,i$, $\, [\bar{X}_{i},\bar{Z}_{j}] = 0 \,\forall \,i\neq j$.
This, then, is a $[[6, 2, 2]]$ stabilizer code.

The plane-filling using the rhombus as a unit is known to ``re-visit" certain segments forming periodically arranged branch-cuts \cite{jain2017nodal,eckhardt1984analytically} (this is not ``tessellation" where there should be uniform and complete filling of the plane). However, this periodic arrangement helps us to identify the edges. Analogous to the surface code \cite{fowler2012surface}, we need to specify the boundaries, the analogues of $X$- and $Z$-edges, which are unions of control-$X$($Z$) edges here, respectively the bold and dashed lines (Fig. \ref{fig:G2}). We define a path by connecting a data vertex of a rhombus to another data vertex of a corresponding copy with respect to the Fundamental Domain of the rhombus.  Each path consists of product of operators which commute with the stabilizers, thus entailing the $Z$($X$) logical operators. 

For instance, the paths containing $X$-ancillae start from the edge $D1-A1$ or $A1-D2$ and terminate at the edge $D5-A4$ or $A4-D6$. The paths for finding $\bar{Z}$ are $D1-A1-D3-A4-D5$, $D1-A1-D3-A4-D6$, $D1-A1-D4-A4-D5$, $\ldots$ where $A's$ are $X$-ancillae. These are the eight paths for the operator $\bar{Z}$. In terms of $Z$, these paths give us eight arrangements of which two are stabilizers ($Z_{1}Z_{3}Z_{5}$ and $Z_{2}Z_{4}Z_{6}$) and the remaining six qualify as logical $\bar{Z}$ operators ($Z_{1}Z_{3}Z_{6}$, $Z_1Z_4Z_5$, $Z_1Z_4Z_6$, $Z_2Z_3Z_5$, $Z_2Z_3Z_6$ and $Z_2Z_4Z_5$). They commute with the stabilizers. In a similar manner, there is another set of paths for finding $\bar{X}$. These are $D1-A2-D3$, $D1-A2-D5$, $D2-A3-D6$, $\ldots$ where $A's$ are $Z$-ancillae. In terms of $X$, these paths give us six possible arrangements ($X_{1}X_{3}$, $X_{1}X_{5}$, $X_{3}X_{5}$, $X_{2}X_{4}$, $X_{2}X_{6}$ and $X_{4}X_{6}$) which commute with the stabilizers. We find that there are two pairs of logical operators satisfying above conditions for logical operators:
$\{\bar{X}_1=X_1X_3$, $\bar{Z}_1=Z_1Z_4Z_6\}$ and $\{\bar{X}_2=X_4X_6$, $\bar{Z}_2=Z_2Z_4Z_5\}$. We may find two different pairs for which logical states  remain the same, thereby producing no new codewords.

Let us consider that errors $\textit{E}_a$ and $\textit{E}_b$ have occurred.  According to the Knill-Laflamme theorem \cite{knill}, the code should be able to distinguish error $\textit{E}_a$ acting on a basis codeword $|\bar{\psi}_i\rangle$ from error $\textit{E}_b$ acting on a different codeword $|\bar{\psi}_j \rangle$, where \{$|\bar{\psi}_i \rangle $($i = 1, 2, \ldots, k$)\}, spans the codeword space. With $E=E_a^\dag E_b$, Knill-Laflamme conditions,
\begin{equation}\label{eq:knill-laflamme}
    \langle\psi_i|E|\psi_j\rangle=C_{ab}\delta_{ij}, \forall |\psi\rangle,
\end{equation}
allow us to ascertain the code distance, where $C_{ab}$ is a Hermitian matrix. The Knill-Laflamme condition is a necessary condition for the code to correct errors $\{E_a\}$. It is also a sufficient condition. Two errors acting on a code may annihilate a codeword ($\bra{\psi_i}E_a^\dagger E_b\ket{\psi_j}=0$) iff $C_{ab}$ does not have a maximum rank. A code for which $C_{ab}$ is (not) singular is (\textit{non})\textit{degenerate} code \cite{gottesman1997stabilizer}. The weight of the shortest $\textit{E}$ in the group containing all stabilizers and codewords for which \eqref{eq:knill-laflamme} does not hold gives the distance of the code. For the code in Fig. \ref{fig:G2}, this weight is found to be two, and hence the distance of the code is two. Therefore, this is a $[[6,2,2]]$ code. It can provide two encryptions and single-qubit error detection but no correction.

\subsection{Genus-two code is not a toric code}
Are toric code and genus-two code same? Is there a map from toric code to genus-two code? We show, by an explicit example, that the spectra of toric code and genus two code are different.

The logical ground state of the unit structure of the $[[6,2,2]]$ (Fig. \ref{fig:G2}) ``genus-two code" is $\ket{0}_{L_1}$:
\begin{alignat}{1}\label{eq:zerol_g2}
    \ket{0}_{L_1}=\frac{1}{\mathcal{N}_1}(\ket{000000}+\ket{001111}+\ket{111100}+\ket{110011}),
\end{alignat}
where $\mathcal{N}_1$ is the normalization factor. Two pairs of logical operators are $\{\bar{X}_1=X_1X_3$, $\bar{Z}_1=Z_1Z_4Z_6\}$ and $\{\bar{X}_2=X_4X_6$, $\bar{Z}_2=Z_2Z_4Z_5\}$.

The set of stabilisers for the unit structure of the toric code $[[5,1,2]]$ (Fig. \ref{fig:reflection}) is $P=\{X_1X_2X_3$, $X_3X_4X_5$, $Z_1Z_3Z_4$, $Z_2Z_3Z_5\}$. For this encoded logical qubit, the logical ground state is $\ket{0}_{L_2}$:
\begin{alignat}{1}\label{eq:zerol_512}
    \ket{0}_{L_2}&=\frac{1}{\mathcal{N}_2}(|00000\rangle+|11100\rangle+|00111\rangle+|11011\rangle).
\end{alignat}
The pair of logical operators is $\{\bar{X}_1=X_1X_4$, $\bar{Z}_1=Z_1Z_2\}$ .

To ``try to map" the logical states, let us apply a CNOT gate between the qubit $4$ and qubit $3$ (qubit $4$ is the control qubit and $3$ is the target qubit) of the $\ket{0}_{L_1}$:

\begin{alignat}{1}\label{l1_cnot}
   CNOT_{4,3}\ket{0}_{L_1}&= \frac{1}{\mathcal{N}_1}(\ket{000000}+\ket{000111}+\ket{110100}+\ket{110011})\nonumber\\
   &=\frac{1}{\mathcal{N}_1}(|00000\rangle+|11100\rangle+|00111\rangle+|11011\rangle)_{1,2,4,5,6}\otimes\ket{0}_3\nonumber\\
   &=\ket{0}_{L_2}\otimes\ket{0}_3.
\end{alignat}

Applying logical $X$ operator ($X_1X_3$) on the transformed state \eqref{l1_cnot}:
\begin{alignat}{1}\label{l2_cnot}
  X_1X_3 (CNOT_{4,3})\ket{0}_{L_1}&=X_1X_3\ket{0}_{L_2}\otimes\ket{0}_3\nonumber\\
  &=\frac{1}{\mathcal{N}_1}(|10000\rangle+|01100\rangle+|10111\rangle+|01011\rangle)_{1,2,4,5,6}\otimes\ket{1}_3\nonumber\\
  &\neq \ket{1}_{L_2}\otimes\ket{1}_3.
\end{alignat}

where $\ket{1}_{L_2}=X_1X_4\ket{0}_{L_2}=\frac{1}{\mathcal{N}_2}(|10010\rangle+|01110\rangle+|10101\rangle+|01001\rangle$).

Even if we may map $\ket{0}_{L_1}$ to $\ket{0}_{L_2}$, the logical operators of [[6,2,2]]-code do not bring $\ket{1}_{L_1}$ to $\ket{1}_{L_2}$. So the spectra of these codes are different.

\section{Planar code structure : Stacking and encryption}\label{sec:stacking}

To construct the planar code using the above unit, we have to make copies by successive reflections about the edges of a unit. The positions of each ancilla and each data qubit are given by specifying the coordinates on the plane. Denoting a point on this planar structure by the coordinates $(p,q)$,  $X$-ancillae are placed at $(\pm 3p,\pm q\sqrt{3})$, $Z$-ancillae are correspondingly placed at coordinates $(\pm 3(p+1/2),\pm\sqrt{3}(q+1/2))$, where $p=0,1,2,\ldots$ and $q=0,1,2,\ldots$

For the positions of data qubits, we have to consider two pairs of coordinates: $(\pm p; p\, {\rm mod}$ $3 \ne 0, \pm q\sqrt{3})$ and $(\pm (2p-1)/2; (2p-1) {\rm mod}$ $3 \ne 0, \pm\sqrt{3}(q-1/2))$ with  $p, q =1,2,\ldots$ 
\subsection{Horizontal stacking}
\begin{figure}[htbp!]
    \begin{center}
    \includegraphics[width=0.88\textwidth]{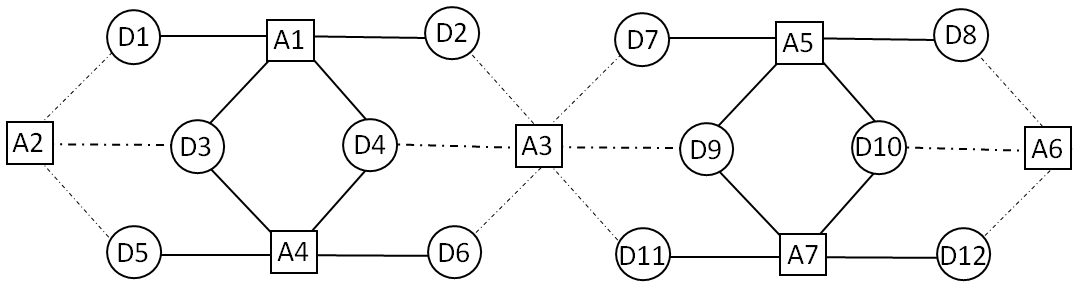}
    \caption{\small{Single unit in Fig. \ref{fig:G2} is stacked horizontally so that we have two copies connected at the $Z$-ancilla, $A_3$. Making another copy along the horizontal direction increases the number of encodings to $k=5$ while keeping the distance, $d=2$. Clearly, this does not aid much to error correction, but it certainly increases the number of encodings substantially.}}
    \label{fig:2horiz}
    \end{center}
\end{figure}

To begin with, let us place a copy of the unit, Fig. \ref{fig:G2}, horizontally, thereby increasing the number of data qubits to $n=12$ and ancilla qubits to $m=7$ (Fig. \ref{fig:2horiz}). The set of stabilizers is $P =\{X_1X_2X_3X_4$, $X_3X_4X_5X_6$, $X_7X_8X_9X_{10}$, $X_9X_{10}X_{11}X_{12}$, $Z_1Z_3Z_5$, $Z_2Z_4Z_6Z_7Z_9Z_{11}$, $Z_8Z_{10}Z_{12}\}$. The directed paths for finding $\bar{Z}$ except for the trivial stabilizer path are - $Z_1Z_3Z_6$, $Z_1Z_4Z_5$, $Z_1Z_4Z_6$, $Z_2Z_3Z_5$, $Z_2Z_3Z_6$, $Z_2Z_4Z_5$, $Z_7Z_9Z_{12}$, $Z_7Z_{10}Z_{11}$, $Z_7Z_{10}Z_{12}$, $Z_8Z_{9}Z_{11}$, $Z_8Z_{9}Z_{12}$ and $Z_8Z_{10}Z_{11}$. The paths for $\bar{X}$ are - $X_1X_3$, $X_1X_5$, $X_3X_5$, $X_2X_4$, $X_2X_6$, $X_4X_6$, $X_7X_9$, $X_7X_{11}$, $X_9X_{11}$, $X_8X_{10}$, $X_8X_{12}$ and $X_{10}X_{12}$. From these possible paths, we construct a complete set of logical operators which commute with the stabilizers and anti-commute pairwise: 
(i) \{$\bar{X}_1=X_2X_6$,  $\bar{Z}_1=Z_1Z_4Z_6$\}, 
(ii) \{$\bar{X}_2=X_4X_6$, $\bar{Z}_2=Z_1Z_4Z_5$\}, 
(iii) \{$\bar{X}_3=X_7X_{11}$, $\bar{Z}_3=Z_8Z_9Z_{11}$\},  
(iv) \{$\bar{X}_4=X_9X_{11}$, $\bar{Z}_4=Z_8Z_9Z_{12}$\}, 
(v) \{$\bar{X}_5=X_2X_7$, $ \bar{Z}_5=Z_2Z_4Z_6$\}. The corresponding logical states constitute the logical codespace, which is denoted by $\mathcal{C}$.

With the number of codewords $k=5$, this is a $[[12,5,2]]$ code, where $k=5$ is actually $n-m=12-7$. In the space $\mathcal{C}$, the minimum weight of error $E$, and hence the code distance turns out to be $d=2$, in accordance with \eqref{eq:knill-laflamme}. The total number of encodings, $k$, is equal to the difference between the number of data qubits, $n$ and ancilla qubits, $m$. In $[[6,2,2]]$ code, we have $k=n-m=6-4=2$ and in $[[12,5,2]]$ code, we have $k=n-m=12-7=5$.
\subsection{Vertical stacking}
\begin{figure}[ht]
    \begin{center}
    \includegraphics[width=0.52\textwidth]{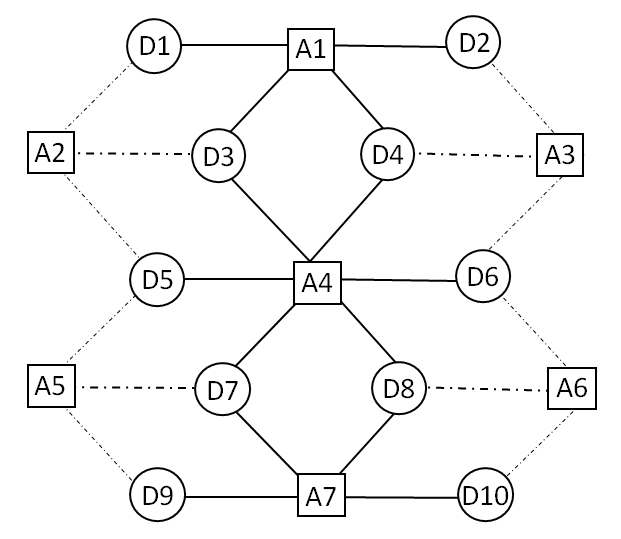}
    \caption{\small{The same unit in Fig. \ref{fig:G2} has been copied vertically. This increases the number of encryptions by only one, giving a total of three logical operators, but the distance has now become three, providing us the possibility of single-qubit error correction.}}
    \label{fig:2vertical}
    \end{center}
\end{figure}

Instead, if we stack a unit vertically on the single unit, Fig. \ref{fig:G2}, the number of physical qubits $n=10$, while the number of ancilla qubits $m=7$, (Fig. \ref{fig:2vertical}). The stabilizers are simply written, viz., $P=\{X_1X_2X_3X_4$, $X_3X_4X_5X_6X_7X_8$, $X_7X_8X_9X_{10}$, $Z_1Z_3Z_5$, $Z_2Z_4Z_6$, $Z_5Z_7Z_9$, $Z_6Z_8Z_{10}\}$. Following the arguments presented above for identifying paths, we obtain  $\bar{X}$ and $\bar{Z}$; the complete set of logical operators commuting with the stabilizers and anti-commuting pairwise is thus 
(i) \{$\bar{X}_1=X_2X_6X_8$, $\bar{Z}_1=Z_1Z_4Z_8Z_9$\}, 
(ii) \{$\bar{X}_2=X_2X_6X_{10}$, $\bar{Z}_2=Z_5Z_7Z_{10}$\}, 
(iii) \{$\bar{X}_3=X_4X_6X_8$, $\bar{Z}_3=Z_2Z_3Z_6$\}.
With these logical operators, the minimum weight of errors for which \eqref{eq:knill-laflamme} does not hold is three. So the code distance is three, constructing a $[[10,3,3]]$ code.

\subsection{Stacking the plane}
\begin{figure}[htbp!]
    \begin{center}
    \includegraphics[width=0.85\textwidth]{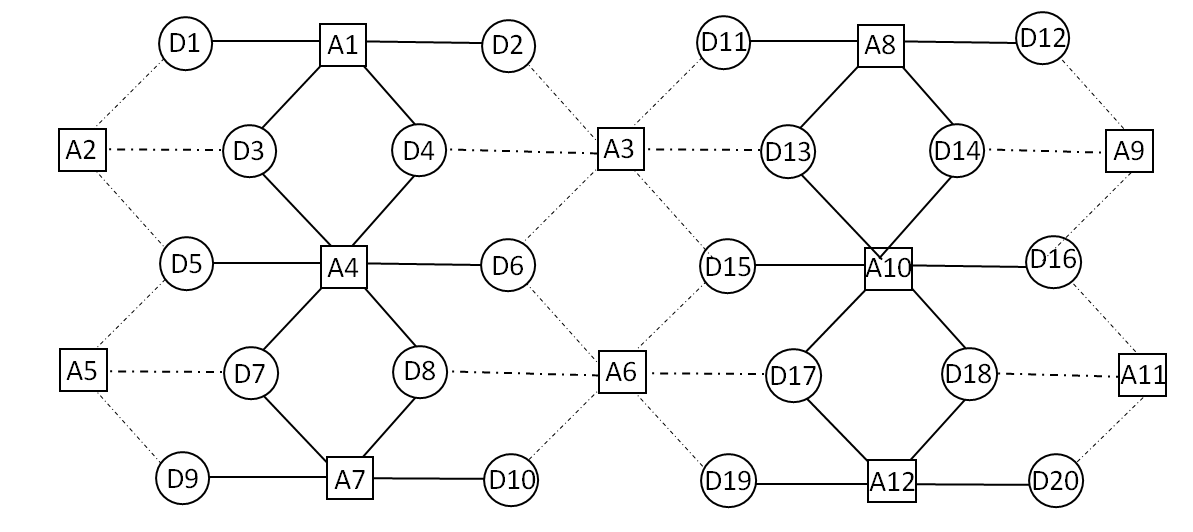}
    \caption{\small {Filling the whole plane in horizontal as well as vertical directions, we find that the number of encodings increases significantly to $k=8$. With distance 3, a single-qubit error correction is possible.}}
    \label{fig:4both}
    \end{center}
\end{figure}

To increase the number of encryptions as well as distance, we need to increase the number of units across the plane, vertically as well as horizontally. For this, we have to fill the plane and use more units as shown in Fig. \ref{fig:4both}. There are $n=20$ data qubits and $m=12$ ancillae, with $k=8$ encryptions and a code distance,  $d=3$. The set of stabilizers is $P=\{X_1X_2X_3X_4$, $X_3X_4X_5X_6X_7X_8$, $X_7X_8X_9X_{10}$, $X_{11}X_{12}X_{13}X_{14}$, $X_{13}X_{14}X_{15}X_{16}X_{17}X_{18}$, $X_{17}X_{18}X_{19}X_{20}$, $Z_1Z_3Z_5$, $Z_2Z_4Z_6Z_{11}Z_{13}Z_{15}$, $Z_5Z_7Z_9$, $Z_6Z_8Z_{10}Z_{15}Z_{17}Z_{19}$, $Z_{12}Z_{14}Z_{16}$, $Z_{16}Z_{18}Z_{20}\}$. The set of logical operators is 
(i) \{$\bar{X}_1=X_1X_5X_9$, $\bar{Z}_1=Z_1Z_3Z_7Z_{10}$\}, 
(ii) \{$\bar{X}_2=X_3X_5X_7$, $\bar{Z}_2=Z_1Z_3Z_8Z_{9}$\},
(iii) \{$\bar{X}_3=X_4X_{6}X_{17}$, $\bar{Z}_3=Z_1Z_4Z_8Z_{9}$\}, 
(iv) \{$\bar{X}_4=X_2X_6X_{17}$, $\bar{Z}_4=Z_2Z_3Z_7Z_{10}$\}, 
(v) \{$\bar{X}_5=X_{6}X_{11}X_{19}$, $\bar{Z}_5=Z_{15}Z_{18}Z_{19}$\}, (vi) \{$\bar{X}_6=X_{11}X_{15}X_{19}$, $\bar{Z}_6=Z_{12}Z_{13}Z_{15}$\}, 
(vii) \{$\bar{X}_7=X_{6}X_{11}X_{17}$, $\bar{Z}_7=Z_{11}Z_{13}Z_{18}Z_{19}$\},
(viii) \{$\bar{X}_8=X_{14}X_{16}X_{20}$, $\bar{Z}_8=Z_{12}Z_{13}Z_{16}$\}.
This is a $[[20,8,3]]$ code. To increase the distance more than 3, we can stack sets of two vertical units shown in Fig. \ref{fig:2vertical}, and thereby fill the whole plane, increasing the encryption. For instance, by placing four copies vertically, we could realize a distance of four, allowing at least a two-qubit error correction. Placing six copies vertically would give a distance of five, and so on.

To simultaneously increase the number of encryptions and the distance of the code, we must stack the unit (Fig. \ref{fig:G2}) horizontally as well as vertically. An equal number of vertical and horizontal stacking arranges the unit structures (Fig. \ref{fig:G2}) in equal number of rows and columns. If the number of these rows and columns is $p$ $(1,2,3,\dots)$, then this planar code structure has $p^2$ units. $p=1$ corresponds to the unit structure Fig. \ref{fig:G2} of this planar code and $p=2$ corresponds to the planar code structure Fig. \ref{fig:4both}. To construct this planar code with $p^2$ units, the required number of data qubits, $n=2p(2p+1)$ and the required number of ancilla qubits, $m=2p(p+1)$. The number of logical encryptions, $k=2p^2$. The code distance of this planar code structure is $d=\lfloor \frac{p+2}{2} \rfloor+1$, where $\lfloor \cdot\rfloor$ is the floor function. So the general form of the code is $\left[\left[2p(2p+1), 2p^2,\lfloor \frac{p+2}{2} \rfloor+1\right]\right]$. The encoding rate of this code is $k/n=p/(2p+1)$. For a single unit ($p=1$), encoding rate is $1/3$ and for a large structure, $p\to\infty$, the encoding rate has the maximum value $k/n=1/2$. 

We can also stack the unit vertically and horizontally unequally. Let us consider a single unit as the structure in Fig. \ref{fig:2vertical}, say $R$, where $n=10$, $m=7$ and $d=3$. If we stack $v$ layers vertically, we will have to add $8v$ data qubits and $6v$ ancillae, where $v=0,1, \dots$ If we add $h$ $R$'s horizontally, we will have to add $10h$ data qubits and $5h$ ancillae, where $h=0,1, \dots$ This will create an L-shaped structure. So, for this kind of structure, we have, $n=10+8v+10h$ and $m=7+6v+5h$, so that $k=3+2v+5h$ and $d=v+3$.

Now, if we have to make a $(v+1)\times(h+1)$ type of matrix of $R$'s, we will have to add another $8vh$ data qubits and $4vh$ ancillae. So, we will have $n=10+8v+10h+8vh$, $m=7+6v+5h+4vh$, $k=3+2v+5h+4vh$ and $d=v+3$.

To incorporate a $v' \times h'$ type matrix anywhere in between the L-shaped structure, where $v$ and $h$ include $v'$ and $h'$, these numbers will change as $n=10+8v+10h+8v'h'$ and $m=7+6v+5h+4v'h'$. The distance is still $d=v+3$.

For a $(v+1)\times(h+1)$ type of matrix, the encoding rate $k/n$ is thus $(3+2v+5h+4vh)/(10+8v+10h+8vh)$ which is $\sim 3/10$ for $R$ and $\sim 1/2$ for very large values of $v,h$.

We thus find that the encoding rate, $k/n$, of this planar code structure is always $ \geq$ $0.3$, and for a large structure, it tends to $0.5$. This is one of the major results of our work.

\section{Comparison of code distance in toric and genus-two codes}\label{app:App III}

In the [[5,1,2]] code shown in Fig. \ref{512code}, the code distance is $2$. Logical operators with higher weight are also possible. The paths $D1-A1-D3-A4-D5$ and $D2-A3-D3-A2-D4$ provide the pair of logical operator $\langle\bar{X}=X_2X_3X_4, \bar{Z}=Z_1Z_3Z_5\rangle$ of weight $3$. Both the operators commute with all the stabilizers of the $[[5,1,2]]$ code and anticommute with each other. In this way we achieved a pair of logical operators of weight $3$ and so the code distance could be 3 making it a [[5,1,3]] code instead. But for the states corresponding to these operators, the minimum weight of error for which  Knill-Laflamme conditions \eqref{eq:knill-laflamme} do not hold is $d=2$, indicating that this has to be a distance $2$ code, hence the code is $[[5,1,2]]$. This is well-expected. 

In the genus-two code, logical operators with smaller code distances are also possible. For example, in the $[[10,3,3]]$ code (Fig. \ref{fig:2vertical}), we may take the pair of logical operators as \{$\bar{Z}=Z_1Z_2$, $\bar{X}=X_1X_5$\} which commute with all the stabilizers and satisfy the anti-commutation relation with each other. The minimum weight of errors for which \eqref{eq:knill-laflamme} do not hold is two. So the code distance is two for these logical operators. In this case, however, Knill-laflamme conditions do not hold for two distinct sets of logical operators corresponding to code distance $2$ and $3$. For the purpose of efficient encoding, one would choose the set of operators with a distance $3$. As discussed above, this does not hold for $[[5,1,2]]$ toric code. Once we have constructed the $\ket{0}_L$ using the stabilizers, we write the logical operators to construct $\ket{1}_L$ with the maximum possible code distance. The following pairs of logical operators work: \{$\bar{X}_1=X_2X_6X_8$, $\bar{Z}_1=Z_1Z_4Z_8Z_9$\}, \{$\bar{X}_2=X_2X_6X_{10}$, $\bar{Z}_2=Z_5Z_7Z_{10}$\}, 
and \{$\bar{X}_3=X_4X_6X_8$, $\bar{Z}_3=Z_2Z_3Z_6$\} have code distance $3$, these are the required logical operators of the code shown in Fig. \ref{fig:2vertical}. Thus, it allows single-qubit error correction. This is an outcome of the increased path length between one $X$ ancilla boundary to another, thereby increasing the weight of possible $\bar{X}$ itself.

In the [[6,2,2]] code, we could construct logical operators of weight two - $\{X_1X_3, Z_1Z_2\}$ and $\{X_4X_6, Z_5Z_6\}$. In this case, the minimum weight of errors for which Knill-Laflamme conditions do not hold is still $2$, providing the code distance $d=2$. But we have selected the weight $3$ logical $Z$ operators $\{Z_1Z_4Z_6, Z_2Z_4Z_5\}$ due to the reflection property of the structure. Since it is our aim to achieve higher distance codes, this makes the [[$2p(2p+1),2p^2,\lfloor\frac{p+2}{2}\rfloor+1$]] code more suitable  for achieving higher encryption rates and distances than a [[$2p(2p+1),2p^2,2$]] code.

\section{Robustness to dephasing noise}

Any logical qubit should be robust against dephasing caused by an external noise. Recently, it has been shown \cite{pal} that certain observables formed by code space population and logical operators in the code space help determine the dynamical behavior of logical qubits. We consider a time-dependent external fluctuating magnetic field in $z$-direction, which acts on the qubits individually (or globally), leading to local (or global) dephasing. To estimate its effect, let us consider the logical $|1\rangle_L$:
\begin{alignat}{1}
|1\rangle_L&= X_1X_3 \ket{0}_L\nonumber\\
&=\frac{1}{\mathcal{N}}(\ket{101000}+\ket{100111}+\ket{010100}+\ket{011011}).
\end{alignat}
Initially, the logical state is $\ket{\psi}_L=\cos{\frac{\theta}{2}}\ket{0}_L+e^{\iota\phi}\sin{\frac{\theta}{2}}\ket{1}_L$, where $\theta \, ( \in [0,\pi])$ and $\phi \, (\in [0, 2 \pi])$ are real parameters. The evolution of $|\psi\rangle_L$ gives the logical Bloch sphere coordinates, $X_L$, $Y_L$ and $Z_L$.

To study the effect of global and local dephasing noise on the logical qubit of Fig. \ref{fig:G2}, we write the randomly fluctuating variable $B(t)$, obeying the Gaussian distribution $P(B)$ \cite{pal}. Thus we have,
\begin{alignat}{1}
  \bigg\langle \exp&{\bigg(\pm \iota \int_{0}^{t}B(t^\prime)dt^\prime \bigg)}\bigg\rangle\nonumber\\
   &=\exp{\bigg[-\frac{1}{2}\bigg\langle\bigg(\int_{0}^{t}B(t^\prime) dt^\prime\bigg)^2\bigg\rangle}\bigg]= e^{-\gamma t/2}
\end{alignat}
assuming the stationarity of the auto-correlation function of delta-correlated noise, with  $\gamma=\braket{[B(0)]^2}$. Assuming the global dephasing process by a single fluctuating variable $B(t)$ along the $z$-direction acting on all data qubits, the Hamiltonian representing the effect of noise may be written as $H_G(t) =\frac{1}{2}B(t)\sum_{i=1}^{6}\sigma_{z_i}$. In case of local dephasing, the Hamiltonian reads as: $H_L(t)=\frac{1}{2}\sum_{i=1}^{6}B_i(t) \sigma_{z_i}$. Following \cite{pal}, we analyze the effect of noise on the $N$-qubit system by grouping the physical states by their magnetization, defined as the difference between the number of spins in the state $\ket{0}$, denoted by $n^\prime$, and the remaining in state $\ket{1}$, $N-n^\prime$. The magnetisation is, $m^\prime=2n^\prime-N$. The logical state $\ket{0}_L$ is written as, $\ket{0}_L=\sum_{m^\prime} \sum_{l=1}^{N_{m^\prime}}b_l^{m^\prime}\ket{b}_l^{m^\prime}$. Dephasing noise changes the  state $\ket{\psi}_L$ to another state $\ket{\psi^\prime}$ , thereby making the density matrix of the logical qubit as $\rho^\prime=\int \ket{\psi^\prime}\bra{\psi^\prime} P(B)dB.$

The Bloch coordinates $\mathcal{R}\equiv\{R_X, R_Y, R_Z\}$ in the new state are obtained by evaluating the expectation values of the logical operators in the evolved state, given by $\braket{{\mathcal{R}}}=Tr[\rho^\prime \bar{\mathcal{L}}]$, where $\bar{\mathcal{L}}\equiv\{\bar{X}, \bar{Y}, \bar{Z}\}$ represents the logical Bloch vectors in the initial state, $\ket{\psi}$. Noise causes leakage from the code space, quantified by $\mathcal{P}\equiv\{p_x, p_y, p_z\}=\braket{ \mathcal{\bar{L}} P_c}$ where 
\begin{equation}
P_c=\frac{1}{2^n}\prod_{P_i\in\langle P\rangle}(I^{\otimes n}+P_i).
\end{equation}

For the single unit structure (Fig. \ref{fig:G2}), in the presence of global dephasing noise, the logical Bloch  coordinates and the values quantifying leakage turn out to be
\begin{alignat}{1}\label{eq:global}
\braket{R_X}=&\frac{1}{2}(1+e^{-2\gamma t})\sin{\theta}\cos{\phi}\nonumber\\
\braket{R_Y}=&-\frac{1}{2}(1+e^{-2\gamma t})\sin{\theta}\sin{\phi}\nonumber\\
\braket{R_Z}=&\cos{\theta}\nonumber\\
p_x=&\frac{1}{32}(3+4 e^{-2\gamma t}+e^{-8\gamma t})\sin{\theta} \cos{\phi} \nonumber\\
p_y=&-\frac{1}{32}(3+4 e^{-2\gamma t}+e^{-8\gamma t})\sin{\theta} \sin{\phi}\nonumber\\
p_z=&\frac{1}{64}(1+9\cos{\theta}-4e^{-2\gamma t}(1-\cos\theta)+3e^{-8\gamma t}(1+\cos\theta)).
\end{alignat}
The terms in $P_c$ contribute to a factor of $2^4/2^6$ in $\langle \mathcal{\bar{L}}P_c\rangle$, thus giving a pre-factor, $1/4$ in $\mathcal{P}$.

In the presence of local dephasing noise, we have
\begin{alignat}{1}\label{eq:local}
\braket{R_X}&=e^{-2\gamma t} \sin{\theta}\cos{\phi} \nonumber\\
\braket{R_Y}&=-e^{-2\gamma t} \sin{\theta}\sin{\phi} \nonumber\\
\braket{R_Z}&=\cos{\theta}\nonumber\\
p_x&=\frac{1}{8}(e^{-2\gamma t}+e^{-8\gamma t})\sin{\theta} \cos{\phi} \nonumber\\
p_y&=-\frac{1}{8}(e^{-2\gamma t}+e^{-8\gamma t})\sin{\theta} \sin{\phi} \nonumber\\
p_z&= \frac{1}{16}(1+3 e^{-8\gamma t})\cos{\theta}.
\end{alignat}

In the case of global dephasing, the off-diagonal terms in the density matrix have coefficients, $c_{\Delta m}=\exp{\left[-\frac{1}{2}\left(\frac{\Delta m}{2}\right)^2\gamma t\right]}$, viz., $1$, $e^{-2\gamma t}$ and $e^{-8\gamma t}$ from Eq. (\ref{eq:global}). These correspond to the terms of form $\ket{b}_l^m\bra{b}_{l'}^{m}$, $\ket{b}_l^m\bra{b}_{l'}^{m\pm4}$ and $\ket{b}_l^m\bra{b}_{l'}^{m\pm8}$, respectively. It may be noted that $\Delta m=0$ quantifies a decoherence free space.

In case of local dephasing, from Eq. (\ref{eq:local}) we see that for $\gamma=0$, the Bloch vector coordinates in the new state, $\ket{\psi'}$ are the same as that in the old state, $\ket{\psi}_L$ and $\mathcal{P}$ has a pre-factor 1/4. The off-diagonal elements, $c_{\Delta n}=\exp{\left[-\frac{\Delta n}{2}\gamma t\right]}$. Comparing with $\exp{[-\gamma t]}$, we infer $\Delta n=2$, an analog of ``Hamming distance" of the code; it is also in agreement with the distance found using the Knill-Laflamme conditions, Eq. \eqref{eq:knill-laflamme}. As $t\to\infty$, $p_x=p_y=0$, which shows that there is no leakage in the $X-Y$ space. Moreover, even in the presence of noise, $\langle R_Z\rangle$ remains unaffected. Thus the code is significantly robust against noise.

A lot of work has been done on stabilizer codes due to their significance in the development of fault-tolerant quantum computation \cite{bonilla2021xzzx,srivastava2022xyz,tomita2014low}. We have presented a code that is inspired by the intricate topologies of the phase space surfaces of quantum billiards \cite{jain2017nodal}. We have given a complete description of the code - from the organization of data and ancillary qubits to the construction of logical operators and entangled states. We have also shown that the code is scalable with a desirably high encoding rate. The design and optimality of the code illustrate beautiful connections between quantum computation, information, geometry, and topology. 
\vskip 0.25 truecm
\noindent
We thank Rhine Samajdar for asking questions leading to Section 3.1.


\begin{thebibliography}{99}
\bibitem{Kitaev} Alexei Kitaev, Ann. Phys. {\bf 303}, 2--30 (2003).

\bibitem{preskill1998fault} John Preskill, {\em Fault-tolerant quantum computation}, in {\em Introduction to quantum computation and information} Eds. H.-K. Lo, S. Popescu, T. Spiller, pp. 213--269 (World Scientific, 1998).

\bibitem{aharonov2008fault} Dorit Aharonov and Michael Ben-Or, SIAM Journal on Computing {\bf 38}, 1207 (2008).

\bibitem{krjj} Komal Kumari, Garima Rajpoot, Sandeep Joshi, and Sudhir R. Jain, Ann. Phys. {\bf 450}, 169222 (2023). 

\bibitem{gottesman1997stabilizer} Daniel Gottesman, {\em Stabilizer codes and quantum error correction}, Ph. D. thesis (California Institute of Technology, 1997).

\bibitem{hamming1986coding} Richard W. Hamming, {\em Coding and information theory}, (Prentice-Hall, Inc., New Jersey, 1986).

\bibitem{huffman2010fundamentals} W. Cary Huffman and Vera Pless, {\em Fundamentals of error-correcting codes}, (Cambridge university press, Inc., Cambridge, 2010).

\bibitem{bravyi1998quantum} Sergey B. Bravyi and A. Yu. Kitaev, arXiv preprint quant-ph/9811052 (1998).

\bibitem{fowler2012surface} Austin G. Fowler, Matteo Mariantoni, John M. Martinis and Andrew N. Cleland, Phys. Rev. A {\bf 86}, 032324 (2012).

\bibitem{chubb2021statistical} Christopher T. Chubb and Steven T. Flammia,  Annales de l’Institut Henri Poincar{\'e} D {\bf 8}, 269--321 (2021).

\bibitem{hutter2014efficient} Adrian Hutter, James R. Wootton and Daniel Loss, Phys. Rev. A {\bf 89}, 022326 (2014).

\bibitem{bombin2007topological} Hector Bombin and Miguel-Angel Martin-Delgado, Phys. Rev. Lett. {\bf 98}, 160502 (2007).

\bibitem{horsman2012surface} Clare Horsman, Austin G. Fowler, Simon Devitt and Rodney Van Meter, New Journal of Physics {\bf 14}, 123011 (2012).

\bibitem{tomita2014low} Yu Tomita and Krysta M. Svore, Phys. Rev. A {\bf 90}, 062320 (2014).

\bibitem{bonilla2021xzzx} J. Pablo Bonilla Ataides, David K. Tuckett, Stephen D. Bartlett, Steven T Flammia and Benjamin J. Brown, Nat. comm. {\bf 12}, 1--12 (2021).

\bibitem{srivastava2022xyz} Basudha Srivastava, Anton Frisk Kockum and Mats Granath, Quantum {\bf 6}, 698 (2022).

\bibitem{weyl1926nachtrag} Hermann Weyl, Mathematische Zeitschrift {\bf 24}, 789--791 (1926).

\bibitem{cartan1927geometrie} {\'E}lie Cartan, Annali di Matematica pura ed applicata {\bf 4}, 209--256 (1927).

\bibitem{coxeter1973regular} Harold Scott Macdonald Coxeter, {\em Regular polytopes}, (Courier Corporation, 1973).

\bibitem{jain2017nodal} Sudhir R. Jain and Rhine Samajdar, Rev. of Mod. Phys. {\bf 89}, 045005 (2017).

\bibitem{eckhardt1984analytically} Bruno Eckhardt, Joseph Ford and Franco Vivaldi, Physica D: Nonlinear Phenomena {\bf 13}, 339--356 (1984).

\bibitem{jain1992periodic} Sudhir R. Jain and H. D. Parab, Journal of Physics A: Mathematical and General {\bf 25}, 6669 (1992).

\bibitem{richens1981pseudointegrable} P. J. Richens and M. V. Berry, Physica D: Nonlinear Phenomena {\bf 2}, 495--512 (1981).

\bibitem{footnote} In the completely different context of Dynamical Systems, there appear invariant integral phase space surfaces for integrable classical billiards which are topologically equivalent to $n-$ tori, thanks to the Liouville-Arnol'd theorem \cite{Arnold}. 

\bibitem{Arnold} V. I. Arnold, {\em Mathematical Methods in Classical Mechanics}, (Springer, Heidelberg, 1989).

\bibitem{footnote1} $\mathcal{G}_1\equiv\{\pm \rm{I}, \pm \iota\rm{I}, \pm X, \pm\iota X, \pm Y, \pm \iota Y, \pm Z, \pm\iota Z\}$ is the set of matrices that forms a group under the operation of matrix multiplication on a single qubit.

\bibitem{knill} Emanuel Knill and Raymond Laflamme, Phys. Rev. A {\bf 55}, 900--911, (1997).

\bibitem{pal} Amit Kumar Pal, Philipp Schindler, Alexander Erhard, {\'A}ngel Rivas, Miguel A. Martin-Delgado, Rainer Blatt, Thomas Monz and Markus P. M{\"u}ller, Quantum {\bf 6}, 632 (2022).




\end{thebibliography}
\end{document}